\documentclass[final,english]{bullsrsl}

\usepackage[latin1]{inputenc}
\usepackage[T1]{fontenc}
\usepackage{natbib} 
\usepackage{graphicx}
\usepackage{hyperref}
\usepackage{aas_macros}

\begin{document}

\title{Modelling the Point Spread Function of ILMT images}

\author[affil={1,2}, corresponding]{Naveen}{Dukiya}
\author[affil={3,1}]{Bhavya}{Ailawadhi}
\author[affil={4,5}]{Talat}{Akhunov}
\author[affil={1,2}]{Monalisa}{Dubey}
\author[affil={6}]{Sara}{Filali}
\author[affil={7,8}]{Paul}{Hickson}
\author[affil={1}]{Priyanshi}{Kumari}
\author[affil={1}]{Gokul Singh}{Mehra}
\author[affil={1}]{Kuntal}{Misra}
\author[affil={9}]{Vibhore}{Negi}
\author[affil={1,10}]{Kumar}{Pranshu}
\author[affil={6}]{Jean}{Surdej}
\author[affil={1,10}]{Sarvesh Kumar}{Yadav}
\author[affil={2}]{Archana}{Gupta}

\affiliation[1]{Aryabhatta Research Institute of Observational Sciences (ARIES), Manora Peak, Nainital-263002, India}

\affiliation[2]{Mahatma Jyotiba Phule Rohilkhand University, Pilibhit Bypass Road, Bareilly 243006, Uttar Pradesh, India}

\affiliation[3]{Physical Research Laboratory, Navrangpura, Ahmedabad, Gujarat-380009, India}

\affiliation[4]{National University of Uzbekistan, Department of Astronomy and Astrophysics, 100174 Tashkent, Uzbekistan}

\affiliation[5]{Ulugh Beg Astronomical Institute of the Uzbek Academy of Sciences, Astronomicheskaya 33, 100052 Tashkent, Uzbekistan}

\affiliation[6]{Institute of Astrophysics and Geophysics, Li\`ege University, All\'ee du 6 Ao\^ut 19c, 4000 Li\`ege, Belgium}

\affiliation[7]{Department of Physics and Astronomy, The University of British Columbia, 6224 Agricultural Road, Vancouver, BC V6T 1Z1, Canada}

\affiliation[8]{Outer Space Institute, The University of British Columbia, 325-6224 Agricultural Road, Vancouver, BC V6T 1Z1, Canada}

\affiliation[9]{Kavli Institute for Astronomy and Astrophysics, Peking University, Beijing 100871, China}

\affiliation[10]{Department of Applied Optics and Photonics, University of Calcutta, Kolkata, 700106, India}

\correspondance{ndookia@gmail.com}

\date{}
\maketitle

\begin{abstract}
We have devised an approach to model the empirical Point Spread Function (PSF) of the images taken with the 4-meter International Liquid Mirror Telescope (ILMT). The characteristics of the ILMT PSF are influenced by the nature of the Liquid Mirror, as well as by instrument specific and atmospheric factors. As the telescope uses Time Delay Integration (TDI) mode for tracking the passing objects, the PSF varies along the declination axis due to differential motion and star-trail curvature. The 5-element optical corrector homogenizes the PSF to some degree, but does not completely eliminate the disturbing effects mentioned above. The PSF in the RA direction is changing due to the sky conditions. We show that we can successfully model the PSF of 4k$\times$4k chunks using PSFEx, accounting for second-order X-Y variability. Our software robustly reproduces the PSF and its variation across the field.
% We compare the PSF photometry for a set of standard stars of an ILMT frame to their cataloged magnitude in order to demonstrate the capabilities and precision of the PSF modeling technique \KM{the last line can be modified based on what is being presented in this paper}. \ND{Done}
\end{abstract}


\section{Introduction} \label{sec:introduction}
The International Liquid Mirror Telescope (ILMT) is a 4m telescope consisting of a rotating dish filled with liquid mercury. The rotating dish forms a parabolic surface that serves as the telescope's primary mirror \citep{surdej_ilmt_2025}. The primary reflected beam is passed through a 5-lens optical corrector to reduce optical aberrations and to straighten the hyperbolic stellar trajectories. It continuously scans a 22.3$'$ wide strip of sky centered at the declination of $29^{o}21'41''$. The CCD operates in drift-scan mode, matching the readout time to the projected transverse velocity of the stars in the CCD plane. The drift-scanning results in an effective exposure time of $\sim 102$s for each source.

As with all optical telescopes, the ILMT also experiences variation in the PSF across its field-of-view (FoV), the details of which are discussed in \autoref{sec:psf_characteristics}. When the variation is significant, it must be properly modeled to obtain accurate photometry from these images. Two common approaches are -- (i) 
% Model the variation as a polynomial in the X \& Y pixel coordinates of the image. -- 
The parameters of the modeling function can be a polynomial function of X \& Y pixels, i.e., $f(\vec{x}, \theta(X, Y))$. -- 
(ii) Break the image into smaller grids within which the PSF is assumed to be constant. Depending on the image dimensions and PSF variability, both approaches have their pros and cons. The details of the variability modeling approach for ILMT images are discussed in \autoref{sec:pipeline}. Conclusions and further discussion are summarized in \autoref{sec:conclusion}. 

\section{PSF Characteristics of the ILMT} \label{sec:psf_characteristics}
Because the ILMT differs from traditional telescopes in its optics and object tracking, its PSF exhibits unique properties. The radial profile of the PSF of ILMT, and Liquid Mirror telescopes in general, consists of 2 parts -- (i) A core limited by seeing and instrumental effects, and -- (ii) A diffused halo caused by scattering from concentric and spiral waves, and from the mylar film that covers the rotating dish \citep{borra_1992_lmt_shot_test, Hickson_lmt_psf_2007}. The diffused halo accounts for $\sim$40\% of the total flux of a stellar source; however, this statistic varies significantly with seeing (Dukiya et al. 2026, under preparation).

Another unique aspect of the ILMT PSF is the change in PSF shape across the detector due to the drift-scan tracking mode. The drift rate of a star varies slightly across the declination direction, and this effect causes variable elongation in the E-W direction. Additionally, the star trail curvature causes variable elongation in the N-S direction. \cite{Hickson_curvature_corrector} designed a 5-element optical corrector that introduces field distortion to counter these effects. The optical corrector attempts to optimize and homogenize the PSF by reducing distortions, but cannot completely eliminate them. Any variability in the E-W direction is averaged out as the source moves across the detector, but the residual distortions cause variability in the N-S direction.

An additional effect is atmospheric seeing and transparency variations as the sources are recorded by the CCD. This effect causes variation in the PSF in the E-W direction. A detailed analysis of the PSF shape and variability will be presented in Dukiya et al. (2026, under preparation).

\section{Modeling the PSF Variability} \label{sec:pipeline}
\begin{figure}[ht]
    \centering
    \includegraphics[width=0.8\linewidth, trim={0 0 8cm 0}, clip]{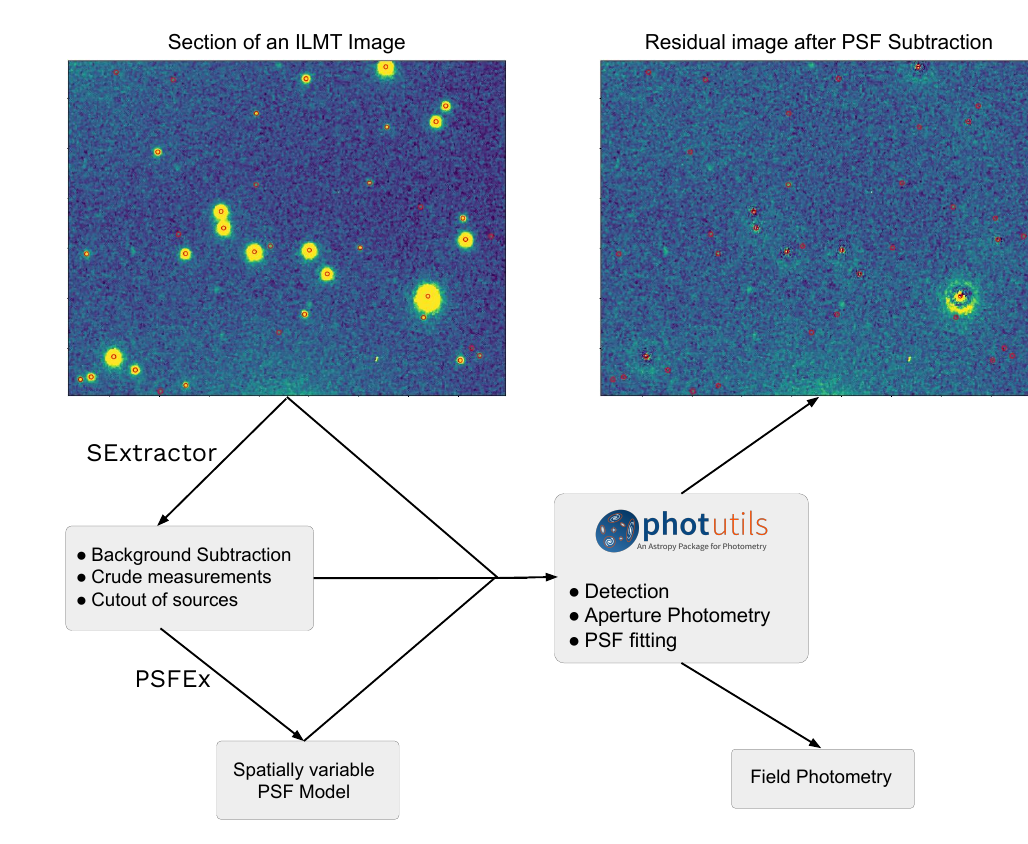}
    \begin{minipage}{12cm}
    \caption{This figure shows the high-level functioning of the ILMT photometry pipeline. The red circles indicate the detected sources.}
    \label{fig:flowchart}
    \end{minipage}
\end{figure}

The images taken using ILMT have the shape 4k$\times$36k pixels (corresponding to 22$'\times201'$). And seeing and transparency changes can cause rapid variations in the PSF in the RA direction, which are difficult to model over 201$'$. Furthermore, processing the entire image in one go proves to be a very computationally intensive task. One solution to address the variability is to divide the image into smaller chunks (say 2k$\times$2k pixels). Even then, the chunks show significant PSF variation inside them, and the edge shared between consecutive chunks will disagree on the PSF shape. Therefore, we adopt a hybrid of the two approaches mentioned in \autoref{sec:introduction}. We divide the entire image into 4k$\times$4k chunks, with slight overlap. For each chunk, we then model the variability in the PSF as second degree polynomials of $X$ and $Y$ pixels using \texttt{SExtractor} + \texttt{PSFEx} \citep{sextractor, sex_psfex_morphometry, psfex_ascl}. The high-level flow of PSF modeling and photometry is shown in \autoref{fig:flowchart}. A 4k$\times$4k chunk of an ILMT image is processed by \texttt{SExtractor}, which creates cutouts of sources, to be used by \texttt{PSFEx} to model the PSF. The PSF model is then used by \texttt{photutils} to perform photometry on the image.

\texttt{PSFEx} is able to generate empirical PSFs robustly and efficiently in an automated manner; however, it does not directly work on FITS images.  It takes as input the catalog generated by \texttt{SExtractor} which contains the crude parameters for the stellar sources and their cutouts. Since ILMT images, even under the best seeing conditions, will be adequately sampled at a pixel scale of 0$''$.33/pixel \citep{surdej_ilmt_2025}, we model the PSF on a pixel basis, i.e., each pixel of the PSF image is treated as a free parameter. We model the PSF up to a radius of $6\times \mathrm{FWHM}$, which contains $\sim 90\%$ of the total flux of a source. 
For reference, the FWHM estimated from 3688 individual frames (observed during Oct 2024 -- May 2025), is 
%$7.6^{+3.1}_{-1.7}$px or 
$1.9^{+1.0}_{-0.5}$ arcsec.
It is impractical to compute the PSF for larger radii due to increasing computational complexity. Through experimentation, we determined that a second order polynomial model can adequately capture the variation of the PSF. 

The photometry pipeline is built in \texttt{Python}, wrapping the functionality of these modules. The output of \texttt{PSFEx} is a \texttt{`.psf'} file which contains a constant PSF array (zeroth order), and additional arrays for each of the following coefficients: $x, x^2, y, xy,$ and $y^2$. These arrays must be evaluated and summed to evaluate the PSF at any given $(x, y)$ location. To perform photometry on the image, we use the \texttt{photutils} \citep{larry_bradley_2024_photutils} module. It provides implementations of source detection, background estimation, and aperture photometry algorithms, as well as a framework for PSF fitting. The \texttt{`.psf'} file is loaded in \texttt{photutils} and evaluated at an arbitrary location using a custom Python class. The PSF photometry is then performed using the \texttt{photutils} module. Sources closer than 1$\times$FWHM are fitted simultaneously. We fit only the central 1$\times$FWHM region containing the highest S/N region pixels for a source. All the 4k$\times$4k chunks are processed sequentially, and a catalog containing the aperture and PSF instrumental magnitudes of all the sources, and the residual image after PSF extraction, is generated as an end product (\autoref{fig:flowchart}). We calculate a zeropoint with a set of standard stars for the entire frame. The zeropoint can vary slightly along the RA due to slight changes in PSF across the 4k$\times$4k chunks, and changes in atmospheric conditions. We correct for this variation by fitting a smoothing spline to the zeropoint trend. A detailed description will be presented in Dukiya et al. (2026, under preparation).

\section{Conclusions and Discussion}
We have built a photometry pipeline capable of performing end-to-end PSF photometry on ILMT images, accounting for PSF variations across the field. The pipeline uses \texttt{Python} programming language to wrap the functionality of \texttt{SExtractor}, \texttt{PSFEx}, and \texttt{photutils}, without using any legacy software. 
From \autoref{fig:flowchart}, we see that the residual image is mostly dominated by the background noise, which highlights the accuracy of the PSF modeling and fitting process. However, bright stars leave a halo in the region beyond the PSF truncation radius. This is because we need to truncate the PSF model at some point due to the fact that the number of pixels to be determined in a PSF model scales as $r^2$, where $ r$ is the radial distance from the center of the PSF. We choose this point to be located at $6 \times$ FWHM, well into the diffused halo region. The effect of this residual halo on the photometric accuracy is insignificant, as the same fraction of flux is left out from each source.

\begin{figure}[ht]
    \centering
    \includegraphics[width=0.75\linewidth, trim={0 0 0 0.7cm}, clip]{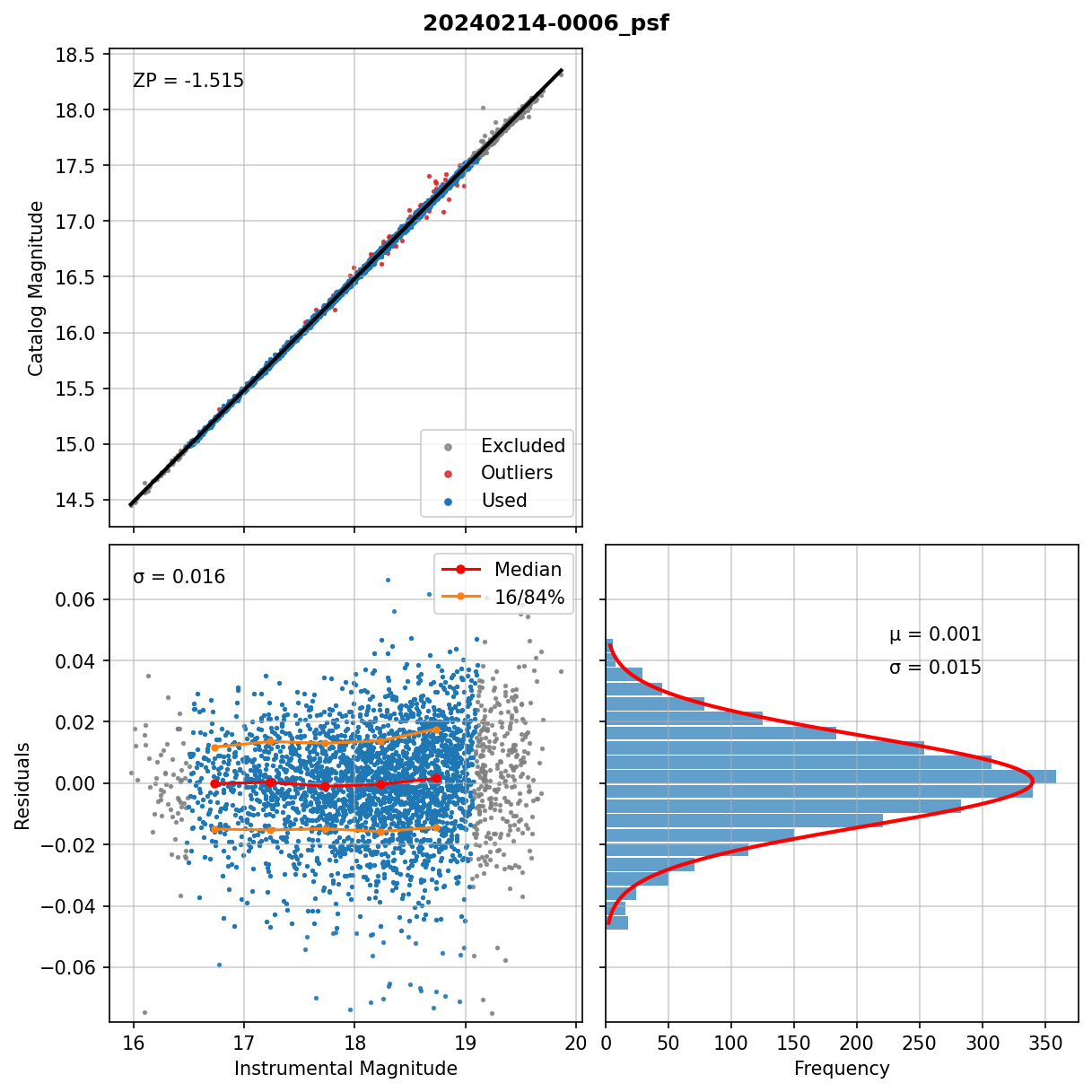}
    \begin{minipage}{12cm}
    \caption{\textit{Top left:} This panel shows the comparison of instrumental PSF magnitudes generated by the ILMT photometry pipeline with their cataloged magnitudes in the PanSTARRS catalog for a set of stars with non-variable magnitudes. This r-band frame was observed on 2024-02-14. \textit{Bottom left:} The residuals after fitting a straight line relation between the instrumental PSF magnitudes and the cataloged magnitudes. The red and orange lines represent the magnitude binned 50$^{th}$ and 16$^{th}$ \& 84$^{th}$ percentiles of the residual scatter, respectively. \textit{bottom right:} The histogram of the residuals fitted with a Gaussian.}
    \label{fig:calibration}
    \end{minipage}
\end{figure}

\autoref{fig:calibration} is a diagnostic plot produced by the pipeline showing the comparison of instrumental PSF magnitudes measured from an \textit{r}-band frame taken on 2024-02-14, with their cataloged magnitudes in the PanSTARRS catalog \citep{panstarrs_survey} for a set of standard sources. The top panel shows a straight-line relation between the two kinds of magnitudes, as expected, with the intercept representing the zeropoint of the frame. The bottom left panel shows the residuals after removing the fitted straight line, and the scatter represents the disagreement between the instrumental and cataloged magnitudes. The magnitude binned 50$^{th}$ and 16$^{th}$ \& 84$^{th}$ percentile lines represent the median deviation and the $\pm 1 \sigma$ scatter, respectively. The bottom right panel shows the histogram of the residuals. The residuals follow a normal distribution, as expected for random, uncorrelated uncertainties. The magnitude of the scatter in the residuals is 0.016 (or 0.015 for the fitted Gaussian) for sources lying in the $\sim 16.5$ to $\sim 19$ mag range. This shows both the precision of PSF modeling and the subsequent photometry and capabilities of the ILMT under good seeing conditions. In case of worse seeing conditions and bad image quality, this scatter tends to increase, reflecting the degraded photometric accuracy.

\label{sec:conclusion}
 
\begin{acknowledgments}
We thank the anonymous referee for their careful review of the manuscript and for providing us with valuable suggestions and scientific insights that enhanced the quality of the paper.
The 4-m International Liquid Mirror Telescope (ILMT) project results from a collaboration between the Institute of Astrophysics and Geophysics (University of Li\`{e}ge, Belgium), the Universities of British Columbia, Laval, Montreal, Toronto, Victoria, and York University, and Aryabhatta Research Institute of Observational Sciences (ARIES, India). The team acknowledges the contributions of ARIES's technical and scientific staff. 
We acknowledge the Belgo-Indian Network for Astronomy and astrophysics (BINA) consortium approved by the International Division, Department of Science and Technology (DST, Govt. of India; DST/INT/BELG/P-09/2017) and the Belgian Federal Science Policy Office (BELSPO, Govt. of Belgium; BL/33/IN12), for allowing us to present our work in the form of a paper in the proceedings of the fourth BINA workshop.
JS wishes to thank Service Public Wallonie, F.R.S.-FNRS (Belgium), and the University of Li\`{e}ge, Belgium, for funding the construction of the ILMT. PH acknowledges financial support from the Natural Sciences and Engineering Research Council of Canada, RGPIN-2019-04369. PH, SF, and JS thank ARIES for hospitality during their visits to Devasthal. We sincerely acknowledge Anna Pospiezalska for her significant contributions to ILMT functioning. MD acknowledges the Innovation in Science Pursuit for Inspired Research (INSPIRE) fellowship award (DST/INSPIRE Fellowship/2020/IF200251) for this work. KP acknowledges the support from the Erasmus+ Programme of the European Union for a research visit to the Institute of Astrophysics and Geophysics, University of Li\`ege, Belgium. TA thanks the Ministry of Higher Education, Science, and Innovations of Uzbekistan (grant FZ-20200929344). JS and KM acknowledge the assistance received from the ANRF SERB-726 VAJRA Faculty Scheme, India. KM acknowledges the support from the BRICS grant DST/ICD/BRICS/Call-5/CoNMuTraMO/2023 (G) funded by the DST, India.
\end{acknowledgments}

\begin{furtherinformation}
\begin{orcids}
\orcid{0000-0002-0394-6745}{Naveen}{Dukiya}
\orcid{0009-0005-3844-3426}{Kumar}{Pranshu}
\orcid{0000-0003-1637-267X}{Paul}{Hickson}
\orcid{0000-0002-7005-1976}
{Kuntal}{Misra}
\orcid{0000-0002-8289-8242}
{Jean}{Surdej}
\end{orcids}

\begin{authorcontributions}
All authors contributed significantly to this work.
\end{authorcontributions}

\begin{conflictsofinterest}
The authors declare no conflict of interest.
\end{conflictsofinterest}
\end{furtherinformation}

\bibliographystyle{bullsrsl-en.bst}
\bibliography{BINA_Naveen}

\end{document}